# Influence of the axial anomaly on the decay $N(1535) \to N\eta$

Lisa Olbrich,[1] Miklós Zétényi,[1,2,3] Francesco Giacosa,[1,4] and Dirk H. Rischke[1,5]

[1]*Institute for Theoretical Physics, Goethe University,*
*Max-von-Laue–Str. 1, D-60438 Frankfurt am Main, Germany*
[2]*Wigner Research Center for Physics, Konkoly Thege Miklós út 29-33, H-1121 Budapest, Hungary*
[3]*ExtreMe Matter Institute EMMI, GSI Helmholtzzentrum für Schwerionenforschung, Darmstadt, Germany*
[4]*Institute of Physics, Jan Kochanowski University, ul. Swietokrzyska 15, 25-406 Kielce, Poland*
[5]*Department of Modern Physics, University of Science and Technology of China, Hefei, Anhui 230026, China*

The decay width of $N(1535) \to N\eta$ is as large as that of $N(1535) \to N\pi$. This is in evident conflict with simple expectations based on flavor symmetry and phase space. Similarly, the decay width of $\Lambda(1670) \to \Lambda(1116)\eta$ is larger than predicted by flavor symmetry. In this work, we propose that the axial $U(1)_A$ anomaly is responsible for an enhanced coupling of (some) excited baryons to the $\eta$ meson. We test this idea by including a new, chirally symmetric but $U(1)_A$ anomalous, term in an effective hadronic model describing baryons and their chiral partners in the mirror assignment. This term enhances the decay of the chiral partners into baryons and an $\eta$ meson, such as $N(1535) \to N\eta$. Moreover, a strong coupling of $N(1535)$ to $N\eta'$ emerges (this is important for studies of $\eta'$ production processes). Our approach shows that $N(1535)$ is predominantly the chiral partner of $N(939)$, and $\Lambda(1670)$ the chiral partner of $\Lambda(1116)$. Finally, our formalism can be used to couple the pseudoscalar glueball $\tilde{G}$ to baryons. We expect a large cross section for the reaction $\bar{p}p \to \tilde{G} \to \bar{p}p(1535)$, which can be experimentally tested in the future PANDA experiment.

## I. INTRODUCTION

The experimental decay width of $N(1535) \to N\eta$ is surprisingly large, $\Gamma_{N(1535) \to N\eta} \simeq (65 \pm 25)$ MeV [1]. In particular, it is as large as the decay width of $N(1535) \to N\pi$, $\Gamma_{N(1535) \to N\pi} = (67.5 \pm 19)$ MeV [1]. On the other hand, flavor symmetry predicts

$$\frac{\Gamma_{N(1535) \to N\eta}}{\Gamma_{N(1535) \to N\pi}} \approx \frac{1}{3} \cos^2 \theta_P \approx 0.17 \; , \qquad (1)$$

where the factor 3 takes into account the pion triplet and $\theta_P \simeq -44.6°$ [2] is the pseudoscalar mixing angle defined by

$$|\eta\rangle = \cos\theta_P \left|\eta_N \equiv (\bar{u}u + \bar{d}d)/\sqrt{2}\right\rangle + \sin\theta_P \left|\eta_S \equiv \bar{s}s\right\rangle \; .$$

This evident violation of flavor symmetry is hard to understand. [Note that phase space would even further reduce the ratio in Eq. (1).] One can easily extend these flavor-symmetry considerations to the whole baryon octet $\{N(1535), \Lambda(1670), \Sigma(1620), \Xi(?)\}$, which decays into the ground-state baryons $\{N(939), \Lambda(1116), \Sigma(1193), \Xi(1338)\}$ and one pseudoscalar meson (for details, see Sec. II). Quite remarkably, while decays involving pions and kaons are well described, also the decay $\Lambda(1670) \to \Lambda(1116)\eta$ is underestimated by arguments based on flavor symmetry. On the contrary, when repeating the study by using the baryon octet $\{N(1650), \Lambda(1800), \Sigma(1750), \Xi(?)\}$, no underestimation of decays involving $\eta$ mesons is found. (A more elaborate discussion of this argument is given in Sec. II of the present paper using a simple model based on flavor symmetry.)

Evidently, the $N(1535)$ must couple to the $\eta$ meson much more strongly than predicted by flavor symmetry, but it is not yet understood how such a strong coupling arises. In some works [3–6], it has been proposed that $N(1535)$ contains a sizable $\bar{s}s$ admixture. Namely, $N(1535)$ arises as a dynamically generated quasi-bound state in the $K\Lambda$ and $K\Sigma$ channels. In other works, see e.g. Refs. [7–9], a pentaquark component of the type $udu\bar{s}s$ is assumed to be present in the resonance $N(1535)$. In both scenarios, $N(1535)$ would be a five-quark object: an enhanced coupling to $\bar{s}s$, and hence to $\eta$ and $\eta'$, emerges naturally in this case. However, in disagreement with these results, recent calculations [10] based on the lattice discretization of quantum chromodynamics (QCD) found that $N(1535)$ has a dominant three-quark core. In Ref. [11] the state N(1535) was studied on the lattice by assuming a three-quark substructure. The resulting axial coupling constant is in agreement with the nonrelativistic quark model, thus also supporting the picture of a large three-quark contribution to the wave function of this resonance. In Ref. [12], the large $N\eta$ branching ratio of some baryon resonances is explained within the constituent quark model in combination with a fine-structure interaction between the quarks in terms of Goldstone-boson exchange.

Another line of research makes use of effective models of QCD based on the linear realization of chiral symmetry. Investigations of nucleon-meson interactions within two-flavor chiral effective models [13] (based on chiral symmetry and the mirror assignment for $N(1535)$ as the chiral partner of the nucleon) showed that the decay width of $N(1535) \to N\eta$ cannot be correctly described. This is indeed expected, because flavor symmetry holds in any chiral model, hence Eq. (1) follows. The inclusion of four baryon multiplets and their mixing was considered in Ref. [14], but the problem with the decay $N(1535) \to N\eta$ is not resolved. For all solutions of the $\chi^2$ fit found there $N(1535) \to N\eta$ is smaller than 10 MeV, hence definitely too small. The small theoretical value turned out to be stable under parameter variations.

Summarizing, the present status shows that further studies are needed to understand the resonance $N(1535)$ and its interaction with the $\eta$ meson. There is, however, an important QCD phenomenon which was not yet systematically taken into account within the above mentioned framework of effective models with linearly realized chiral symmetry, but can provide an answer to the above problem: the axial $U(1)_A$ anomaly. [In the chiral limit, the QCD Lagrangian possesses a $U(1)_A$ symmetry, which is broken by quantum fluctuations [15]]. The axial anomaly is known to be responsible for the mass difference of $\eta$ and $\eta'$ compared to pions and kaons, respectively. Note that the axial anomaly is suppressed in the large-$N_c$ expansion [16], but is known to be exceptionally large: in contrast to other large-$N_c$ suppressed terms, the chiral anomaly is typically not negligible for $N_c = 3$.

Then, the large decay width of $N(1535)$ to $N\eta$ is intuitively explained as follows: quantum fluctuations related to the anomaly couple $N(1535)$ to $N$ via emission of two gluons in the isoscalar-pseudoscalar channel $I = 0$, $J^{PC} = 0^{-+}$. Since this di-gluon couples with the same intensity to the quark-antiquark pairs $\bar{u}u$, $\bar{d}d$, and $\bar{s}s$, it couples almost exclusively to $\eta$ and $\eta'$ and is thus responsible for a decay width which is enhanced compared to simple flavor-symmetry arguments. [For a recent study of the effect of the anomaly on various mesons, see Ref. [17].]

Yet, the technical question is how to achieve such a coupling without spoiling chiral symmetry $SU(3)_L \times SU(3)_R$. The answer is that the desired anomaly term can be easily constructed when the previously mentioned mirror assignment is taken into account [18]. In the two-flavor version of the mirror assignment, it is possible to construct a chirally invariant mass term involving the nucleon and its chiral partner. This mass term and various generalizations of it have been at the basis of many works in the vacuum [13, 14, 19–21] and at non-vanishing density [22–26]. Just as for the chiral mass term, it is possible to build an analogous pseudoscalar term which couples the nucleon and its chiral partner preserving $SU(2)_L \times SU(2)_R$. This new term, however, breaks $U(1)_A$ and yields a coupling to the mesons $\eta$ and $\eta'$. When interpreting $N(1535)$ as the chiral partner of the nucleon, an enhanced decay $N(1535) \to N\eta$ can be obtained by adjusting the respective coupling constant. We present this line of arguments in Sec. III A. While the treatment is completely general, for specific results we use the two-flavor version of the so-called extended Linear Sigma Model (eLSM) developed in Refs. [13, 21].

When considering the three-flavor version of the mirror model, one can write analogous anomalous terms (Sec. III B). Within this framework, one has four baryonic multiplets from the very beginning. As a specific model, we shall use the three-flavor version of the eLSM developed in Ref. [14]. Interestingly, the axial anomaly naturally provides some peculiar mutual interaction between baryons which helps us to identify the octet $\{N(1535)$, $\Lambda(1670)$, $\Sigma(1620)$, $\Xi(?)\}$ as the chiral partners of the ground-state baryons. The numerical implications of the axial anomaly are then studied in Sec. III B 3.

As a last step, we use the developed mathematical structure to couple the pseudoscalar glueball to baryons (Sec. IV). Although at present no data exist, because the pseudoscalar glueball has not been discovered yet, one can still draw conclusions which can be of use when more experimental information will be available. In particular, we shall find that this pseudoscalar glueball can be observed in the reaction $p\bar{p} \to p\bar{p}(1535)$, which can be studied at the future PANDA experiment [27]. Conclusions and an outlook are presented in Sec. V. Some details of the calculations are relegated to the appendices.

Clarifying the role of the axial anomaly is not only relevant for vacuum spectroscopy. There are at least two related fields where such investigations are of interest: (i) Understanding the coupling of baryons to the $\eta$ (as well as $\eta'$) meson is important for the study of mesonic nuclei, where an $\eta$ meson is bound to an atomic nucleus. As discussed in Refs. [28–30], these studies allow to test the axial anomaly and its purported change in the medium. Namely, measurements of such bound states in nuclei are a direct probe of axial-singlet dynamics. Ongoing experiments at COSY, as well as new experiments at ELSA and GSI/FAIR try to find such bound states [31]. (ii) Neutron stars represent an excellent laboratory for hadronic matter under extreme conditions. The development of a three-flavor chiral model, which correctly describes the axial anomaly, can help to understand the role of hyperons in neutron stars [26, 32].

## II. MODEL BASED ON FLAVOR SYMMETRY

Spontaneous breaking of the chiral symmetry leads to a residual $U(N_f)_V$ flavor symmetry (for $N_f$ degenerate quark flavors). It is therefore instructive to first consider a simple model based on this symmetry alone (i.e., without the full chiral symmetry and without terms parametrizing the axial anomaly) and to study the decays of a baryon resonance with negative parity into a ground-state baryon and a pseudoscalar meson.

Let us first define the fields corresponding to the decay products that we aim to study. The nonet of pseudoscalar states $\{\boldsymbol{\pi}, K^\pm, K^0, \bar{K}^0, \eta(547), \eta'(958)\}$ is contained in the $3 \times 3$ matrix $P$:

$$P = \frac{1}{\sqrt{2}} \begin{pmatrix} \frac{\eta_N + \pi^0}{\sqrt{2}} & \pi^+ & K^+ \\ \pi^- & \frac{\eta_N - \pi^0}{\sqrt{2}} & K^0 \\ K^- & \bar{K}^0 & \eta_S \end{pmatrix}.$$

The fields $\eta_N$ and $\eta_S$ are related to the physical fields $\eta = \eta(547)$ and $\eta' = \eta'(958)$ by a standard $O(2)$ rotation,

$$\begin{pmatrix} \eta \\ \eta' \end{pmatrix} = \begin{pmatrix} \cos\theta_P & \sin\theta_P \\ -\sin\theta_P & \cos\theta_P \end{pmatrix} \begin{pmatrix} \eta_N = (\bar{u}u + \bar{d}d)/\sqrt{2} \\ \eta_S = \bar{s}s \end{pmatrix}, \quad (2)$$

where $\theta_P \simeq -44.6°$ [2]. [Using a different value, such as $-42°$ found in Ref. [33], would lead only to minor changes of our results.] The ground-state positive-parity octet of baryons $\{N(939), \Lambda(1116), \Sigma(1193), \Xi(1338)\}$ is described by the $3 \times 3$ matrix $O$, and for the excited octet of negative-parity baryons we introduce the matrix $O_*$:

$$O \equiv \begin{pmatrix} \frac{\Lambda}{\sqrt{6}} + \frac{\Sigma^0}{\sqrt{2}} & \Sigma^+ & p \\ \Sigma^- & \frac{\Lambda}{\sqrt{6}} - \frac{\Sigma^0}{\sqrt{2}} & n \\ \Xi^- & \Xi^0 & -\frac{2\Lambda}{\sqrt{6}} \end{pmatrix}, \quad (3)$$

$$O_* \equiv \begin{pmatrix} \frac{\Lambda_*}{\sqrt{6}} + \frac{\Sigma^0_*}{\sqrt{2}} & \Sigma^+_* & p_* \\ \Sigma^-_* & \frac{\Lambda_*}{\sqrt{6}} - \frac{\Sigma^0_*}{\sqrt{2}} & n_* \\ \Xi^-_* & \Xi^0_* & -\frac{2\Lambda_*}{\sqrt{6}} \end{pmatrix}.$$

At a microscopic level, a flavor transformation corresponds to a simple rotation of the underlying quark field,

$$q \to U_V q \,,$$

where $q = (u, d, s)^T$ and $U_V$ is a $3 \times 3$ unitary matrix belonging to the group $U(3)_V$. When applied to the matrices $P$, $O$, and $O_*$, the following transformation behavior emerges:

$$P \to U_V P U_V^\dagger \,, \quad O \to U_V O U_V^\dagger \,, \quad O_* \to U_V O_* U_V^\dagger \,.$$

Under parity, the fields transform as $O \to \gamma^0 O(t, -\mathbf{x})$ and $O_* \to -\gamma^0 O_*(t, -\mathbf{x})$, while under charge conjugation as $O \to C\bar{O}^T$ and $O_* \to -C\bar{O}_*^T$ (where the transposition $T$ acts in Dirac and flavor spaces and $C$ is the charge-conjugation matrix).

It is now easy to construct a flavor, parity, and charge-conjugation invariant model which couples $O_*$ to $O$ and $P$:

$$\mathcal{L}_V = i\lambda_V \text{Tr}(\bar{O}PO_* - \bar{O}_*PO) \,. \quad (4)$$

The coupling constant $\lambda_V$ is dimensionless. The explicit form of the Lagrangian after evaluation of the trace as well as the expressions of the corresponding decay widths are presented in App. A.

The term in Eq. (4) is not the only flavor-invariant term that can be written down. It is, however, the term dominating in the large-$N_c$ expansion, according to which $\lambda_V \propto \sqrt{N_c}$ (it describes a standard decay by creating a quark-antiquark pair from the vacuum). Further terms are given by

$$i\beta_V \text{Tr}(\bar{O}O_*P - \bar{O}_*OP) \,,$$
$$i\gamma_V \text{Tr}(\bar{O}O_* - \bar{O}_*O)\text{Tr}\, P \,. \quad (5)$$

As shown in App. B, these are, however, suppressed in the large $N_c$ limit, $\beta_V \propto 1/\sqrt{N_c}$ and $\gamma_V \propto 1/N_c^{3/2}$. This is due to the fact that they involve (at least) two gluons in the intermediate state. Such terms are typically negligible compared to the dominant one and will be set to zero in the remainder of this section. Yet, as mentioned above, an important exception concerns terms that arise from the axial anomaly. As we shall see, when the anomaly is coupled to the model, a term of the type $\sim \gamma_V$ in Eq. (5) emerges (together with various other terms) from the anomalous coupling of baryons to mesons (for details, see Sec. III). This term can be responsible for the fact that some decay widths are larger than expected.

TABLE I: Results from the flavor model with $O_* \equiv \{N(1535), \Lambda(1670), \Sigma(1620), \Xi(?)\}$.

| | Flavor model [MeV] | Experiment [1] [MeV] |
|---|---|---|
| $\Gamma_{N(1535) \to N\pi}$ | $67.5 \pm 19$ | $67.5 \pm 19$ |
| $\Gamma_{N(1535) \to N\eta}$ | $\mathbf{4.3 \pm {}^{1.3}_{1.1}}$ | $\mathbf{40 - 91}$ |
| $\Gamma_{\Lambda(1670) \to N\bar{K}}$ | $6.0 \pm {}^{1.8}_{1.6}$ | $5 - 15$ |
| $\Gamma_{\Lambda(1670) \to \Sigma\pi}$ | $21.3 \pm {}^{6.4}_{5.6}$ | $6.25 - 27.5$ |
| $\Gamma_{\Lambda(1670) \to \Lambda\eta}$ | $\mathbf{0.6 \pm {}^{1.8}_{1.6}}$ | $\mathbf{2.5 - 12.5}$ |
| $\Gamma_{\Sigma(1620) \to N\bar{K}}$ | $32 \pm {}^{26}_{20}$ | – |
| $\Gamma_{\Sigma(1620) \to \Lambda\pi}$ | $21.7 \pm {}^{6.5}_{5.7}$ | – |
| $\Gamma_{\Sigma(1620) \to \Sigma\pi}$ | $39 \pm {}^{12}_{10}$ | – |
| $\Gamma_{\Sigma(1620) \to \Sigma\eta}$ | kin. not allowed | – |
| $\Gamma_{\Sigma(1560) \to N\bar{K}}$ | $27 \pm {}^{22}_{17}$ | – |
| $\Gamma_{\Sigma(1560) \to \Lambda\pi}$ | $19.8 \pm {}^{6.0}_{5.2}$ | – |
| $\Gamma_{\Sigma(1560) \to \Sigma\pi}$ | $34 \pm {}^{10}_{9.1}$ | – |
| $\Gamma_{\Sigma(1560) \to \Sigma\eta}$ | kin. not allowed | – |

We now turn to numerical results. We shall consider two distinct models. In the first one, the octet $O_*$ describes the baryon states $\{N(1535), \Lambda(1670), \Sigma(1620), \Xi(?)\}$, while in the second one, it represents the heavier states $\{N(1650), \Lambda(1800), \Sigma(1750), \Xi(?)\}$. (Indeed, one could go further: each baryon octet with quantum numbers $J^P = \frac{1}{2}^-$ can be assigned to $O_*$.)

**(1) Model with $O_* \equiv \{N(1535), \Lambda(1670), \Sigma(1620), \Xi(?)\}$**

For the first assignment, the parameter $\lambda_V$ can be determined by fitting the decay width of $N_* \equiv N(1535) \to N\pi$ to the experimentally well determined value $(67.5 \pm 19)$ MeV [1]:

$$\lambda_V = \lambda_V^{N(1535)} = 1.37 \pm 0.19 \,.$$

Using this result, we can compute the remaining decay widths, which are summarized in Tab. I. Most of the decay widths are in agreement with the experimental data. For completeness, in the second part of the table, we also present the results for the assignment $\Sigma^* \equiv \Sigma(1560)$.

There are two important mismatches, both of them linked to the $\eta$ meson, which were mentioned in the introduction as a motivation of the present paper: the decay widths $N(1535) \to N\eta$ and $\Lambda(1670) \to \Lambda\eta$ (bold-faced in Tab. I) come out too small by about one order of magnitude. Evidently, flavor symmetry is not sufficient to describe the decays of $\{N(1535), \Lambda(1670), \Sigma(1620), \Xi(?)\}$ states into a ground-state baryon and an $\eta$ meson. These results show that an additional component is needed when the $\eta$ meson is considered. As we shall see, the chiral anomaly does the desired job in both channels.

TABLE II: Results from the flavor model with $O_* \equiv \{N(1650), \Lambda(1800), \Sigma(1750), \Xi(?)\}$.

|  | Flavor model [MeV] | Experiment [1] [MeV] |
|---|---|---|
| $\Gamma_{N(1650) \to N\pi}$ | $84 \pm 23$ | $84 \pm 23$ |
| $\Gamma_{N(1650) \to N\eta}$ | $8.7 \pm^{2.6}_{2.2}$ | $15.4 - 37.5$ |
| $\Gamma_{N(1650) \to \Lambda K}$ | $13.2 \pm^{3.9}_{3.4}$ | $5.5 - 25.5$ |
| $\Gamma_{\Lambda(1800) \to N\bar{K}}$ | $8.2 \pm^{2.4}_{2.1}$ | $50 - 160$ |
| $\Gamma_{\Lambda(1800) \to \Sigma\pi}$ | $28.2 \pm^{8.2}_{7.2}$ | seen |
| $\Gamma_{\Lambda(1800) \to \Lambda\eta}$ | $3.09 \pm^{0.91}_{0.79}$ | $2 - 44$ |
| $\Gamma_{\Sigma(1750) \to N\bar{K}}$ | $23.0 \pm^{6.7}_{5.9}$ | $6 - 64$ |
| $\Gamma_{\Sigma(1750) \to \Lambda\pi}$ | $28.2 \pm^{8.2}_{7.2}$ | seen |
| $\Gamma_{\Sigma(1750) \to \Sigma\pi}$ | $53 \pm^{16}_{14}$ | $< 12.8$ |
| $\Gamma_{\Sigma(1750) \to \Sigma\eta}$ | $2.37 \pm^{0.69}_{0.60}$ | $9 - 88$ |
| $\Gamma_{\Sigma(1620) \to N\bar{K}}$ | $18.1 \pm^{5.3}_{4.6}$ | − |
| $\Gamma_{\Sigma(1620) \to \Lambda\pi}$ | $24.2 \pm^{7.2}_{6.2}$ | − |
| $\Gamma_{\Sigma(1620) \to \Sigma\pi}$ | $44 \pm^{13}_{11}$ | − |
| $\Gamma_{\Sigma(1620) \to \Sigma\eta}$ | kin. not allow | − |

**(2) Model with $O_* \equiv \{N(1650), \Lambda(1800), \Sigma(1750), \Xi(?)\}$**

For the second assignment, using the decay width of $N(1650) \to N\pi$ to fit the parameter $\lambda_V$, we obtain

$$\lambda_V = \lambda_V^{N(1650)} = 1.45 \pm 0.20 \ .$$

The results for the remaining decay widths are listed in Tab. II. In this case the value for the width of the experimentally well-known decay $\Lambda(1800) \to \Lambda\eta$ is in agreement with the data. The decay width $N(1650) \to N\eta$ comes out too small by (at least) a factor of 1.4 (if we consider the maximum theoretical and minimum experimental values). However, note that the experimental range of the $N\eta$ branching ratio was located at smaller values in the previous edition of the PDG [34] (between 5.5 and 25.5 MeV). In the new edition of 2016 [1], it seems that only the analysis of Ref. [35] has been taken into account, while the much smaller result of Ref. [36] has been neglected (it was quoted in the list of experiments included in the average, but was not used to compute the latter). Quite peculiarly, while the result of Ref. [35] is reported in the table of decay modes, later on no average or fit is reported for this branching ratio. Concerning the decay $\Sigma(1750) \to \Sigma\eta$, the maximum theoretical value of the decay width of 3 MeV underestimates the minimum experimental value of 9 MeV by a factor 3. [The experimental result was originally determined in a single experiment which was performed over four decades ago [37].] In conclusion, at present there is no stringent evidence of enhanced decays with an $\eta$ meson in the final state.

There is, however, a disagreement concerning the decays $\Lambda(1800) \to N\bar{K}$ and $\Sigma(1750) \to \Sigma\pi$. The theoretical decay width for $\Lambda(1800) \to N\bar{K}$ is too small. Interestingly, in a recent partial-wave analysis of $\bar{K}N$ scattering [38], the numerical value reads $\Gamma_{\Lambda(1800) \to N\bar{K}} = (33 \pm 20)$ MeV, which agrees with our theoretical result. Concerning the decay $\Sigma(1750) \to \Sigma\pi$, the PDG [1] quotes only (a rather small) upper limit in its summarizing table. Here, our result is too large. In the analysis of Ref. [38], the decay $\Sigma(1750) \to \Sigma\pi$ is clearly seen. [The result of Ref. [38] is cited by the PDG [1] but is not included in the summary table.] Although the errors are large, the central value reads 58 MeV, in good agreement with our theoretical value. Finally, according to Ref. [38] the decay width for $\Sigma(1750) \to \Lambda\pi$ is of the order of 20 MeV, which is also compatible with our results. In summary, while some of our results are in disagreement with the PDG [1], most of them agree well with the most recent and complete analysis of the decays of Ref. [38].

Summarizing the results of this section, the inability to properly describe the decays into $\eta$ in a flavor-symmetric model calls for an explanation: as we shall see, the inclusion of the $U(1)_A$ anomaly provides a candidate.

## III. ANOMALY TERM IN THE MIRROR ASSIGNMENT

In this section we investigate chiral models for baryons (and in particular the eLSM) and introduce a term which preserves chiral symmetry but explicitly breaks the axial $U(1)_A$ symmetry. We treat baryons in the so-called mirror assignment. We first present the basic ideas for a two-flavor model, where it is easier to see how the mechanism works. Then, we extend our considerations to a three-flavor model. For both cases we discuss the consequences of the anomaly for the decays of baryons.

### A. The case $N_f = 2$

Let us briefly discuss the mirror assignment in the two-flavor case $N_f = 2$, i.e., for the nucleon and its chiral partner, originally introduced in Ref. [18] and further studied in Refs. [13, 19, 21–25] and references therein.

We recall that a quark field $q = (u, d)^T$ is split into its left- and right-handed components by using the chiral projection operators $P_{L/R} = (1 \mp \gamma_5)/2$:

$$q_{L/R} = P_{L/R} q \ .$$

Under a chiral $SU(2)_L \times SU(2)_R$ transformation, the two components transform differently:

$$q_L \to U_L q_L , \quad q_R \to U_R q_R . \tag{6}$$

Here, $U_L \in SU(2)_L$ and $U_R \in SU(2)_R$ are two, in general distinct, matrices. We now turn to composite baryon fields. In the mirror assignment, one starts with two nucleon fields, $\Psi_1$ and $\Psi_2$. These two fields mix (see below) to form the nucleon $N = (p, n)^T$, where $p$ describes the proton and $n$ the neutron, and its chiral partner, which for the sake of definiteness we assume to be the resonance $N(1535)$. The central point is how the baryon fields transform under chiral transformations. In the mirror assignment, the following transformations are postulated:

$$\Psi_{1,L} \to U_L \Psi_{1,L} , \quad \Psi_{1,R} \to U_R \Psi_{1,R} ,$$
$$\Psi_{2,L} \to U_R \Psi_{2,L} , \quad \Psi_{2,R} \to U_L \Psi_{2,R} ,$$

where $\Psi_{k,L/R} = P_{L/R}\Psi_k$. One observes that the baryon field $\Psi_1$ transforms under chiral transformations just as the underlying fundamental quark field $q$. The left-handed part transforms under $U_L$ and the right-handed part under $U_R$. However, this is not the case for the baryon field $\Psi_2$, which transforms in a mirror way: the left-handed part transforms under $U_R$ and vice versa. We also recall that under parity transformations the fields behave as $\Psi_1 \to \gamma^0 \Psi_1(t, -\mathbf{x})$ and $\Psi_2 \to -\gamma^0 \Psi_2(t, -\mathbf{x})$, while under charge-conjugation transformations as $\Psi_1 \to C\bar{\Psi}_1^T$ and $\Psi_2 \to -C\bar{\Psi}_2^T$.

The peculiar mirror transformation allows to introduce a chirally (as well as $P$ and $C$) invariant mass term:

$$\begin{aligned}\mathcal{L}_{m_0}^{N_f=2} &= m_0 \left( \bar{\Psi}_2 \gamma^5 \Psi_1 - \bar{\Psi}_1 \gamma^5 \Psi_2 \right) \\ &= m_0(\bar{\Psi}_{2,L}\Psi_{1,R} - \bar{\Psi}_{2,R}\Psi_{1,L} \\ &\quad + \bar{\Psi}_{1,R}\Psi_{2,L} - \bar{\Psi}_{1,L}\Psi_{2,R}) ,\end{aligned} \tag{7}$$

which was first written down in Ref. [18] and further investigated in various works, e.g. Refs. [13, 22, 23, 25, 39] and references therein. The physical fields $N$ and $N_*$ corresponding to the nucleon and to $N(1535)$ are given by:

$$\begin{pmatrix} N \\ N_* \end{pmatrix} = \frac{1}{\sqrt{2\cosh\delta}} \begin{pmatrix} e^{\delta/2} & \gamma_5 e^{-\delta/2} \\ \gamma_5 e^{-\delta/2} & -e^{\delta/2} \end{pmatrix} \begin{pmatrix} \Psi_1 \\ \Psi_2 \end{pmatrix} , \tag{8}$$

where

$$\cosh\delta = \frac{m_N + m_{N_*}}{2m_0} . \tag{9}$$

For $m_0 \to 0$ one has $\delta \to \infty$, and the mixing vanishes: $N = \Psi_1$, $N_* = -\Psi_2$. The numerical value of $m_0$ depends on the model employed. For instance, in Ref. [21] the value $m_0 = (459 \pm 117)$ MeV has been obtained. In other works, it ranges between 200 and 700 MeV [22–25]. Interestingly, the Lagrangian can be made dilatation-invariant via the substitution

$$m_0 \to a\chi + bG$$

where $\chi$ is a four-quark field [pre-dominantly corresponding to $f_0(500)$] [26, 40], while $G$ is a dilaton field [pre-dominantly corresponding to $f_0(1710)$] [41]. The previously introduced constant $m_0$ is obtained after condensation of $\chi$ and $G$: $m_0 = a\chi_0 + bG_0$. For further study of this term, see Refs. [23, 24, 26, 42]. We now turn to the axial anomaly. First, we notice that the combination

$$\begin{aligned}&\bar{\Psi}_2\Psi_1 - \bar{\Psi}_1\Psi_2 \\ &= \bar{\Psi}_{2,L}\Psi_{1,R} + \bar{\Psi}_{2,R}\Psi_{1,L} - \bar{\Psi}_{1,R}\Psi_{2,L} - \bar{\Psi}_{1,L}\Psi_{2,R}\end{aligned}$$

is chirally invariant, but has negative parity (charge conjugation is positive). As it stands, it cannot be a term of an effective Lagrangian. However, when including mesons, things change. For $N_f = 2$, scalar and pseudoscalar mesons are incorporated in the field

$$\Phi = S + iP \tag{10}$$
$$= \frac{1}{\sqrt{2}} \begin{pmatrix} \frac{\sigma_N + a_0^0}{\sqrt{2}} & a_0^+ \\ a_0^- & \frac{\sigma_N - a_0^0}{\sqrt{2}} \end{pmatrix} + \frac{i}{\sqrt{2}} \begin{pmatrix} \frac{\eta_N + \pi^0}{\sqrt{2}} & \pi^+ \\ \pi^- & \frac{\eta_N - \pi^0}{\sqrt{2}} \end{pmatrix} ,$$

where $\boldsymbol{\pi}$ is the pion field, $\eta_N$ is the non-strange two-flavor version of the $\eta$ (and $\eta'$) meson(s) [see Eq. (2) in the previous section], $\boldsymbol{a}_0$ is identified with $a_0(1450)$, and $\sigma_N$ with $f_0(1370)$ [2, 43]. Under chiral transformations of the underlying quark fields,

$$\Phi \to U_L \Phi U_R^\dagger . \tag{11}$$

Moreover, one has $\Phi \to \Phi^\dagger$ under parity and $\Phi \to \Phi^T$ under charge-conjugation transformations. Hence, the negative-parity term

$$\det\Phi - \det\Phi^\dagger$$

is invariant under chiral $SU(2)_L \times SU(2)_R$ transformations, but not under $U(1)_A$. Namely,

$$\det\Phi \to (\det U_L)(\det\Phi)(\det U_R)^* ,$$

which is clearly invariant under $SU(2)_L \times SU(2)_R$, since in this case $\det U_L = \det U_R = 1$. On the other hand, an axial $U(1)_A$ transformation corresponds to the choice $U_L = e^{i\alpha} = U_R^\dagger$, which implies that

$$\det\Phi \to e^{4i\alpha}\det\Phi \neq \det\Phi . \tag{12}$$

This is why terms involving the determinant are usually employed to model the axial anomaly.

It is now possible to construct a parity-even chiral invariant which couples baryons to mesons in the following way:

$$\mathcal{L}_A^{N_f=2} = \lambda_A^{N_f=2}(\det\Phi - \det\Phi^\dagger)(\bar{\Psi}_2\Psi_1 - \bar{\Psi}_1\Psi_2) , \tag{13}$$

where the parameter $\lambda_A$ has dimension [energy$^{-1}$]. The previous equation contains the main idea of the present work. After spontaneous symmetry breaking, the scalar field $\sigma_N$ acquires a nonzero expectation value: $\langle\sigma_N\rangle =$ $\phi_N$. After the shift $\sigma_N \to \sigma_N + \phi_N$ and in the absence of (axial-)vector mesons, one has

$$\det \Phi - \det \Phi^\dagger = -i\left[(\sigma_N + \phi_N)\eta_N - \boldsymbol{a_0} \cdot \boldsymbol{\pi}\right] \ .$$

One observes that a direct coupling of the meson $\eta_N$ to the baryonic combination $\bar{\Psi}_2\Psi_1 - \bar{\Psi}_1\Psi_2$ appears.

When the hadronic model contains (axial-)vector mesons as well [such as in the eLSM for $N_f = 2$ [13, 21, 43]], then $\phi_N = Z_\pi f_\pi$, where $f_\pi = 92.1$ MeV is the pion decay constant and $Z_\pi \simeq 1.79$. In addition, in order to ensure canonically normalized kinetic terms, one also has to replace $\boldsymbol{\pi} \to Z_\pi \boldsymbol{\pi}$ and $\eta_N \to Z_{\eta_N} \eta_N$, where $Z_{\eta_N} \simeq Z_\pi$. These changes only quantitatively influence our picture.

Summarizing, in the two-flavor version of the eLSM [43] the term in the Lagrangian describing the chiral anomaly for baryons as a function of the physical fields reads:

$$\mathcal{L}_A^{N_f=2} = -\frac{i\lambda_A^{N_f=2}}{\cosh \delta}[Z_{\eta_N}(\sigma_N + \phi_N)\eta_N - Z_\pi \boldsymbol{a_0} \cdot \boldsymbol{\pi}]$$
$$\times (\bar{N}\gamma_5 N + \bar{N}_*\gamma_5 N_* - \sinh\delta \bar{N}N_* + \sinh\delta \bar{N}_*N) \ .$$

One observes that a contribution to the decay width of $N_* \to N\eta$ arises:

$$i\lambda_A^{N_f=2} \tanh\delta \phi_N Z_{\eta_N} \bar{N}\eta_N N_* + \text{ h.c. }.$$

In Ref. [13] it is shown that within the eLSM with baryons for two flavors the decay width of $N_* \to N\eta$ turns out to be far too small if we choose $N(1535)$ as the chiral partner of the nucleon. Interestingly, the contribution of the anomaly term solves this problem. In App. C we report the details of the calculation as well as the numerical results of the eLSM for $N_f = 2$. However, it must be also stressed that the case $N_f = 2$, even if interesting because it shows that a new decay mechanism is possible, does not allow to make additional predictions. For that purpose, we turn to the case $N_f = 3$ in the next subsection.

As a further technical remark, we point out that it is also possible to write an anomalous term

$$(\det \Phi + \det \Phi^\dagger)(\bar{\Psi}_2\gamma^5\Psi_1 - \bar{\Psi}_1\gamma^5\Psi_2) \ ,$$

that represents a possible further anomalous contribution to $m_0$. After condensation one obtains a contribution to $m_0$ proportional to $\phi_N^2$.

A similar approach for coupling the baryons to the mesons $\eta$ and $\eta'$ via the QCD axial anomaly can be followed to study the QED anomaly. First, one has to replace $\det \Phi - \det \Phi^\dagger$ with $F_{\mu\nu}\tilde{F}^{\mu\nu}$, where $F_{\mu\nu} = \partial_\mu A_\nu - \partial_\nu A_\mu$ is the electromagnetic field tensor and $\tilde{F}_{\mu\nu} = \frac{1}{2}\varepsilon_{\mu\nu\alpha\beta}F^{\alpha\beta}$ its dual. Second, one has to take into account that $\psi_1 = (p_1, n_1)^T$ and $\psi_2 = (p_2, n_2)^T$ as well as the different electric charges of the quark emitting two gluons, leading to the Lagrangian

$$\mathcal{L}_{A,QED}^{N_f=2} = \lambda_{A,QED}^{N_f=2} e^2 F_{\mu\nu}\tilde{F}^{\mu\nu}$$
$$\times \left[\frac{4}{9}(\bar{p}_2 p_1 - \bar{p}_1 p_2) + \frac{1}{9}(\bar{n}_2 n_1 - \bar{n}_1 n_2)\right] \ .$$

Various interaction terms emerge, some of which lead to decays of the type $n(1535) \to n\gamma\gamma$ and $p(1535) \to p\gamma\gamma$. At present, there is no data for such reactions. Moreover, these decays can also take place via other processes involving the QCD and QED anomalies, leading to the transition chains $N(1535) \to NX \to N\gamma\gamma$ with $X = \pi^0, \eta, \eta'$. We recall that $\pi^0, \eta, \eta'$ couple to $\gamma\gamma$ via the QED anomaly, $XF_{\mu\nu}\tilde{F}^{\mu\nu}$, see e.g. Ref. [44] for a description of such processes within a linear sigma model.

### B. The case $N_f = 3$

In the three-flavor case, both mesons and baryons are described by $3 \times 3$ matrices (in nonet and octet matrix fields, respectively). Moreover, in Ref. [14] it was shown that, using the quark-diquark picture and requiring a mirror assignment, for $N_f = 3$ one can construct four baryonic multiplets. In the following, we briefly recall their main characteristics, then we concentrate on implications for decays involving the $\eta$ meson.

#### 1. The baryonic fields

The fundamental chiral transformation for the three-flavor case has the same form as in Eq. (6), with $q = (u, d, s)^T$ and unitary $3 \times 3$ matrices $U_{L,R}$. For hadronic fields composed of quarks, the chiral transformation for mesonic fields is a straightforward generalization of Eq. (6), while for baryonic fields it is less obvious. The mesonic field matrix $\Phi$ containing scalar and pseudoscalar degrees of freedom reads

$$\Phi = S + iP = \frac{1}{\sqrt{2}}\begin{pmatrix} \frac{\sigma_N + a_0^0}{\sqrt{2}} & a_0^+ & K_0^{*+} \\ a_0^- & \frac{\sigma_N - a_0^0}{\sqrt{2}} & K_0^{*0} \\ K_0^{*-} & \bar{K}_0^{*0} & \sigma_S \end{pmatrix} + \frac{i}{\sqrt{2}}\begin{pmatrix} \frac{\eta_N + \pi^0}{\sqrt{2}} & \pi^+ & K^+ \\ \pi^- & \frac{\eta_N - \pi^0}{\sqrt{2}} & K^0 \\ K^- & \bar{K}^0 & \eta_S \end{pmatrix} \ , \qquad (14)$$

where the identification of the non-strange fields is identical to the $N_f = 2$ case. In addition, $\sigma_S$ corresponds predominantly to $f_0(1500)$ [with an admixture of $f_0(1710)$, which is, however, predominantly gluonic; for more details, see Ref. [41]], the field $K_0^*$ corresponds to $K_0^*(1430)$, and $K$ to the kaons. As already stated in Sec. II, see Eq. (2), $\eta_N$ and $\eta_S$ mix and generate the physical states $\eta$ and $\eta'$. The chiral transformation of $\Phi$ is the same as in Eq. (11). Under $SU(3)_L \times SU(3)_R \times U(1)_A$ the determinant transforms as

$$\det \Phi \to e^{6i\alpha} \det \Phi \neq \det \Phi \ , \qquad (15)$$

i.e., as in Eq. (12), one observes an explicit breaking of $U(1)_A$. The anomalous terms in the Lagrangian are given by $\det \Phi + \det \Phi^\dagger$, see Refs. [43, 45] and references therein, and by $(\det \Phi - \det \Phi^\dagger)^2$, see Ref. [2] and references therein. They generate an additional mass difference between $\eta$ and $\pi$, as discussed in Ref. [28].

TABLE III: Parity and charge-conjugation transformations of the baryonic fields.

| Field | Parity | Charge conjugation |
|---|---|---|
| $N_{1R}$ | $-\gamma^0 N_{2L}(t,-\boldsymbol{x})$ | $-i\gamma^2 (N_{2L})^\star$ |
| $N_{1L}$ | $-\gamma^0 N_{2R}(t,-\boldsymbol{x})$ | $-i\gamma^2 (N_{2R})^\star$ |
| $N_{2R}$ | $-\gamma^0 N_{1L}(t,-\boldsymbol{x})$ | $-i\gamma^2 (N_{1L})^\star$ |
| $N_{2L}$ | $-\gamma^0 N_{1R}(t,-\boldsymbol{x})$ | $-i\gamma^2 (N_{1R})^\star$ |
| $M_{1R}$ | $-\gamma^0 M_{2L}(t,-\boldsymbol{x})$ | $i\gamma^2 (M_{2L})^\star$ |
| $M_{1L}$ | $-\gamma^0 M_{2R}(t,-\boldsymbol{x})$ | $i\gamma^2 (M_{2R})^\star$ |
| $M_{2R}$ | $-\gamma^0 M_{1L}(t,-\boldsymbol{x})$ | $i\gamma^2 (M_{1L})^\star$ |
| $M_{2L}$ | $-\gamma^0 M_{1R}(t,-\boldsymbol{x})$ | $i\gamma^2 (M_{1R})^\star$ |

In Ref. [14], four baryonic multiplets were constructed from the quark-diquark picture. They transform under chiral transformations as follows:

$$\begin{aligned}
N_{1R} &\to U_R N_{1R} U_R^\dagger \ , & N_{1L} &\to U_L N_{1L} U_R^\dagger \ , \\
N_{2R} &\to U_R N_{2R} U_L^\dagger \ , & N_{2L} &\to U_L N_{2L} U_L^\dagger \ , \\
M_{1R} &\to U_L M_{1R} U_R^\dagger \ , & M_{1L} &\to U_R M_{1L} U_R^\dagger \ , \\
M_{2R} &\to U_L M_{2R} U_L^\dagger \ , & M_{2L} &\to U_R M_{2L} U_L^\dagger \ .
\end{aligned} \qquad (16)$$

The chiral transformation matrix acting from the left acts on the quark, while that on the right acts on the diquark. As one observes, $N_1$ and $N_2$ transform (as far as transformation from the left is concerned) in the standard way, while $M_1$ and $M_2$ transform (from the left) in a mirror way. These fields behave under parity and charge-conjugation transformations as shown in Tab. III. Baryonic fields with definite behavior under parity transformations are introduced as:

$$B_N = \frac{N_1 - N_2}{\sqrt{2}} \ , \ B_{N_*} = \frac{N_1 + N_2}{\sqrt{2}} \ ,$$
$$B_M = \frac{M_1 - M_2}{\sqrt{2}} \ , \ B_{M_*} = \frac{M_1 + M_2}{\sqrt{2}} \ ,$$

where now $B_N$ and $B_M$ have positive parity and $B_{N_*}$ and $B_{M_*}$ have negative parity. In Ref. [14] it was shown that $N(1535)$ is always the chiral partner of the nucleon, but depending on the values of the coupling constants of the underlying Lagrangian, $N(1535)$ can be (predominantly) a state of the multiplet $B_{M_*}$ or of the multiplet $B_{N_*}$. Both possibilities give a similarly good description of masses and decay widths (with the notable exception of the decay $N(1535) \to N\eta$). In this work we shall restrict ourselves to the former possibility, for reasons which will become apparent below. Thus, in the following the negative-parity mirror field $B_{M_*}$ is regarded as the chiral partner of the ground-state baryon field $B_N$, while $B_{N_*}$ is the chiral partner of $B_M$. Taking the two-flavor limit, $B_N \to \Psi_1$ and $B_{M_*} \to \Psi_2$. While in principle mixing between $B_N$, $B_M$, $B_{N_*}$, and $B_{M_*}$ takes place, see Ref. [14], in order to keep the discussion simple we will neglect this for the remainder of this paper (a detailed study of mixing should, nevertheless, be subject of future work). We thus simply identify:

$$\begin{aligned}
B_N &\equiv \{N(939), \Lambda(1116), \Sigma(1193), \Xi(1318)\} \ , & (17) \\
B_M &\equiv \{N(1440), \Lambda(1600), \Sigma(1660), \Xi(1690)\} \ , & (18) \\
B_{M_*} &\equiv \{N(1535), \Lambda(1670), \Sigma(1620), \Xi(?)\} \ , & (19) \\
B_{N_*} &\equiv \{N(1650), \Lambda(1800), \Sigma(1750), \Xi(?)\} \ , & (20)
\end{aligned}$$

see Eq. (3) for the matrix form. For the full eLSM Lagrangian for $N_f = 3$ we refer to Ref. [14] [and for the mesonic sector to Ref. [2]].

### 2. The Lagrangian

In terms of the fields $N_1$, $N_2$, $M_1$, and $M_2$ the Lagrangian describing the chiral anomaly is constructed as:

$$\begin{aligned}
\mathcal{L}_A^{N_f=3} =& \lambda_{A1}(\det \Phi - \det \Phi^\dagger)\mathrm{Tr}(\bar{M}_{1R} N_{1L} - \bar{N}_{1L} M_{1R} \\
& - \bar{M}_{2L} N_{2R} + \bar{N}_{2R} M_{2L}) \\
& + \lambda_{A2}(\det \Phi - \det \Phi^\dagger)\mathrm{Tr}(\bar{M}_{1L} N_{1R} - \bar{N}_{1R} M_{1L} \\
& - \bar{M}_{2R} N_{2L} + \bar{N}_{2L} M_{2R}) \ , \qquad (21)
\end{aligned}$$

where the parameters $\lambda_{A1}$ and $\lambda_{A2}$ have dimension [energy$^{-2}$]. This term is analogous to the two-flavor version in Eq. (13). Using Eq. (16), it is easy to show that chiral invariance under $SU(3)_R \times SU(3)_L$ is fulfilled, but due to the determinant, $U(1)_A$ is explicitly broken [see Eq. (15)].

In terms of the fields with definite parity $B_N$, $B_M$, $B_{N_*}$, and $B_{M_*}$ [which, in the limit of zero mixing, are assigned as in Eqs. (17) – (20)], the Lagrangian takes the form:

$$\mathcal{L}_A^{N_f=3} = \frac{\lambda_{A1} + \lambda_{A2}}{2}(\det\Phi - \det\Phi^\dagger)\text{Tr}(\bar{B}_{M_*}B_N - \bar{B}_N B_{M_*} - \bar{B}_{N_*}B_M + \bar{B}_M B_{N_*})$$
$$- \frac{\lambda_{A1} - \lambda_{A2}}{2}(\det\Phi - \det\Phi^\dagger)\text{Tr}(\bar{B}_N \gamma_5 B_M + \bar{B}_M \gamma_5 B_N + \bar{B}_{N_*}\gamma_5 B_{M_*} + \bar{B}_{M_*}\gamma_5 B_{N_*}) \,, \quad (22)$$

where the $\gamma_5$ matrix originates from writing out the chiral projection operators. This Lagrangian is analogous to the two-flavor version of Eq. (13) upon setting $\Psi_1 = B_N$, $\Psi_2 = B_{M_*}$ and identifying:

$$\lambda_A^{N_f=3} = \frac{\lambda_{A1} + \lambda_{A2}}{2} \,.$$

The first line of Eq. (22) shows that the anomaly gives contributions to terms which couple mesons to $B_N$ and $B_{M_*}$ (as well as $B_M$ and $B_{N_*}$), which allows for the possibility of an enhanced decay of the type

$$B_{M_*} \to B_N \eta \,.$$

If we identify $B_{M_*} \equiv \{N(1535), \Lambda(1670), \Sigma(1620), \Xi(?)\}$ as the chiral partners of the ground-state baryons, we are then naturally lead to the possibility that the anomalous terms can give rise to an enhanced decay of $N(1535)$ into $N\eta$. In contrast, the anomaly does not produce terms where mesons couple to $B_N$ and $B_{N_*}$ or $B_M$ and $B_{M_*}$. Identifying $B_{N_*} \equiv \{N(1650), \Lambda(1800), \Sigma(1750), \Xi(?)\}$, it is then obvious that there is no additional contribution from the anomaly to the decay of $N(1650)$ into $N\eta$.

While the fit of Ref. [14] to masses and decay widths in principle also allows for the possibility to identify $B_{N_*}$ with $\{N(1535), \Lambda(1670), \Sigma(1620), \Xi(?)\}$, the anomaly would then not give an enhanced decay width of $N(1535)$ into $N\eta$. For this reason we discarded this option from the very beginning of our discussion. Vice versa, requiring a proper description of the decay width $N(1535) \to N\eta$ via the anomaly forces us to discard the scenario where $B_{N_*} \equiv \{N(1535), \Lambda(1670), \Sigma(1620), \Xi(?)\}$.

In conclusion, just as we have seen in Sec. II, the anomaly could explain decays involving the $\eta$ mesons, which are enhanced above the values predicted by flavor symmetry alone. We finally note that the second line of Eq. (22) describes interactions of the (pseudo)scalar mesons with two baryons of equal parity. However, decays involving the $\eta$ meson are kinematically forbidden.

The present discussion of the anomaly shows that $N(1535), \Lambda(1670), \Sigma(1620), \Xi(?)$ are predominantly the chiral partners of the ground-state baryons. However, even if the states $N(1535)$ and $N$ are (predominantly) chiral partners, there is no mass degeneracy (indeed, the ratio $\frac{M_{N(1535)} - M_{N(1940)}}{2(M_{N(1535)} + M_{N(1940)})} \sim 50\%$ shows a large effect of spontaneous symmetry breaking) and the interaction of $N(1535)$ with $N\pi$ is not small, as the corresponding decay rate shows. Similar considerations hold for the other members of the multiplet. The situation is expected to be different for heavier baryons, where a mass degeneracy between chiral partners and weak interactions with pions (as well as kaons and the $\eta_8$ meson) are expected [46–48]. Our chiral model takes these features into account in a natural way: namely, when the chiral condensate $\phi_N \to 0$, one recovers the degeneracy of $N(1535)$ and $N$ and the decay $N(1535) \to N\pi$ vanishes. Thus, we expect that such a model is suitable to perform a detailed study of $p\bar{p}$ scattering in the future. [For a first study in this direction, see Ref. [49], where nucleon-nucleon scattering close to threshold is studied within the eLSM. A study of $pp \to ppX$ where $X = \omega, \rho, ...$, is currently ongoing. Another interesting application of $p\bar{p}$ scattering is connected to the search for a putative pseudoscalar glueball, see Sec. V.] Interestingly, as discussed in Ref. [47], models based on chiral doublets can be useful to understand massive baryons. Heavy baryons can be described by a larger value of $m_0$ (which, in turn, implies a minor role of spontaneous symmetry breaking in generating their masses).

### 3. Consequences of the anomaly

We now discuss the consequences of the chiral anomaly. Using the matrix form (14) of the (pseudo)scalar meson field $\Phi$, one has

$$\det\Phi - \det\Phi^\dagger = \frac{i}{2\sqrt{2}}\Big[Z_{\eta_S}\phi_N^2\eta_S + 2Z_{\eta_N}\phi_N\phi_S\eta_N\Big] + \ldots, \quad (23)$$

where spontaneous symmetry breaking has been taking into account via the shifts $\sigma_N \to \phi_N + \sigma_N$ and $\sigma_S \to \phi_S + \sigma_S$. In the following, for all constants we use the numerical values as given in Ref. [2]. For instance, the vacuum expectation values are $\phi_N = 164.6$ MeV and $\phi_S = 126.2$ MeV. Due to the fact that (axial-)vector degrees of freedom are present in the eLSM, also wave-function renormalization factors occur, leading to the field redefinitions $\boldsymbol{\pi} \to Z_\pi\boldsymbol{\pi}$, $\eta_N \to Z_{\eta_N}\eta_N$, $\eta_S \to Z_{\eta_S}\eta_S$, and $K^\pm \to Z_K K^\pm$. Here, $Z_\pi = Z_{\eta_N} = 1.79$, $Z_{\eta_S} = 1.47$, and $Z_K = 1.56$, for details, see Ref. [2].

Equation (23) shows that $\det\Phi - \det\Phi^\dagger$ is proportional to the fields $\eta_S$ and $\eta_N$ (the dots refer to further nonlinear terms in the fields). In the $U(3)_V$-limit (with $\phi_N = \sqrt{2}\phi_S$), one has

$$\det\Phi - \det\Phi^\dagger = \frac{iZ_\pi}{2}\sqrt{\frac{3}{2}}\phi_N^2\eta_0 + \ldots$$

where, as expected, the isosinglet-pseudoscalar combination $\eta_0 = (\sqrt{2}\eta_N + \eta_S)/\sqrt{3}$ enters. Thus, the anomaly term of Eq. (22) causes an enhanced interaction with the flavor-singlet field $\eta_0 = \sqrt{2}\,\mathrm{Tr}\,P$. Note that in terms of physical fields one has:

$$\eta_0 = \sqrt{2}\,\mathrm{Tr}\,P = \frac{\eta}{\sqrt{3}}\left(\sqrt{2}\cos\theta_P + \sin\theta_P\right)$$
$$+ \frac{\eta'}{\sqrt{3}}\left(\cos\theta_P - \sqrt{2}\sin\theta_P\right)\;.$$

Using $\theta_P = -44.6°$ [2], one obtains $\eta_0 = 0.18\eta + 0.98\eta'$.

We now turn to the interaction with baryons and to one particular term which has a nonzero contribution to decays. Restricting ourselves to the fields $B_N$ and $B_{M^*}$, the interaction Lagrangian takes the form

$$\mathcal{L}_A^{N_f=3} = i\tilde{\lambda}_A \mathrm{Tr}(\bar{B}_{M_*}B_N - \bar{B}_N B_{M_*})\mathrm{Tr}\,P + \ldots\;,\quad (24)$$

with

$$\tilde{\lambda}_A = \frac{\lambda_{A1}+\lambda_{A2}}{2}\frac{\sqrt{3}Z_\pi}{2}\phi_N^2\;,$$

where dots refer to interactions with more than one mesonic field and to flavor-breaking terms. Here, we recall that $B_N \equiv \{N(939), \Lambda(1116), \Sigma(1193), \Xi(1338)\}$ and $B_{M_*} \equiv \{N(1535), \Lambda(1670), \Sigma(1620), \Xi(?)\}$. Identifying $B_N \equiv O$ and $B_{M_*} \equiv O_*$ (see Sec. II, first model) we thus recognize that the anomaly yields a term of the form $i\tilde{\lambda}_A \mathrm{Tr}(\bar{O}O_* - \bar{O}_* O)\mathrm{Tr}\,P$, cf. Eq. (5). Hence, by taking into account the axial anomaly, we obtain an improved flavor model:

$$\mathcal{L}_V^{\mathrm{improved}} = \mathcal{L}_V + \mathcal{L}_A^{N_f=3}$$
$$= i\lambda_V \mathrm{Tr}(\bar{O}PO_* - \bar{O}_* PO)$$
$$+ i\tilde{\lambda}_A \mathrm{Tr}(\bar{O}O_* - \bar{O}_* O)\mathrm{Tr}\,P\;.\quad (25)$$

By using the decay widths of $N(1535) \to N\eta$, we obtain the following values of the parameters:

$$\tilde{\lambda}_A = 11 \pm 0.6\;,$$

and

$$\lambda_A^{N_f=3} = \frac{\lambda_{A1}+\lambda_{A2}}{2} = (264 \pm 13)\,\mathrm{GeV}^{-2}\;.$$

Besides the decay width of $N(1535) \to N\eta$, it was not possible to reproduce the decay width of $\Lambda(1670) \to \Lambda\eta$ in a model with flavor symmetry only, compare Tab. I. Now, including the anomaly term (22), the model reproduces this decay width properly:

$$\Gamma_{\Lambda(1670)\to\Lambda\eta} = (8.7 \pm 0.4)\,\mathrm{MeV}\;.$$

According to Ref. [1] the numerical value should lie in the range of $(2.5-12.5)$ MeV, see Tab. I. Hence, the increase caused by the anomaly is in very good agreement with the present experimental value. [Note that there is a second solution, which is realized for $\tilde{\lambda}_A = -19 \pm 0.6$, or $(\lambda_{A1}+\lambda_{A2})/2 = (-451 \pm 13)$ GeV$^{-2}$. However, this solution implies that $\Gamma_{\Lambda(1670)\to\Lambda\eta} = (51\pm4)$ MeV, which is unacceptably large.] Other decays into the $\eta$ meson are kinematically forbidden.

Another interesting consequence of the anomaly is the enhanced coupling of the nucleon $N$ and its chiral partner $N_*$ to $\eta'$. When expanding the Lagrangian $\mathcal{L}_V^{\mathrm{improved}}$ one obtains:

$$\mathcal{L}_V^{\mathrm{improved}} = ig_{\eta NN_*}\eta\left(\bar{N}_*N - \bar{N}N_*\right)$$
$$+ ig_{\eta'NN_*}\eta'\left(\bar{N}_*N - \bar{N}N_*\right)$$
$$+ ig_{\pi NN_*}\boldsymbol{\pi}\cdot\left(\bar{N}_*\boldsymbol{\tau}N - \bar{N}\boldsymbol{\tau}N_*\right) + \ldots\;,$$

where

$$g_{\eta NN_*} = \frac{i\lambda_V}{2}\cos\theta_P + \frac{i\tilde{\lambda}_A}{\sqrt{6}}(\sqrt{2}\cos\theta_P+\sin\theta_P) \simeq 1.9\;,$$

$$g_{\eta'NN_*} = \frac{i\lambda_V}{2}\sin\theta_P + \frac{i\tilde{\lambda}_A}{\sqrt{6}}(-\sqrt{2}\sin\theta_P+\cos\theta_P) \simeq 7.2\;,$$

$$g_{\pi NN_*} \simeq -\frac{\lambda_V}{2} = -0.7\;.$$

When the anomaly is neglected, $\tilde{\lambda}_A = 0$, we have $g_{\eta NN_*} \simeq |g_{\eta'NN_*}| \simeq 0.5$. Hence, our results confirm that the anomaly leads to an increased coupling of $N$ and $N_*$ to $\eta'$. This result is in qualitative agreement with Ref. [50], where $g_{\eta'NN_*} \simeq 3.7$ was found by studying scattering processes of the type $pn \to pn\eta'$. Scattering processes can also be studied within the eLSM, see Ref. [49]. We leave a more detailed discussion of this issue for future work.

## IV. INTERACTIONS OF THE PSEUDOSCALAR GLUEBALL WITH BARYONS

The search for glueballs is an interesting topic in hadronic physics [51–55]. Quenched lattice-QCD calculations [56] predict a rich glueball spectrum. Unquenched lattice-QCD studies confirm these results [57], but the mixing of glueballs with ordinary mesons could not yet be determined. In the future, the ongoing BESIII [58] and, most importantly, the planned PANDA experiments [27] can shed light on these missing states of QCD. In the framework of the eLSM, glueballs were studied in Refs. [41, 59–63]. In particular, the pseudoscalar glueball is directly linked to the chiral anomaly [60, 61]. The mathematical formalism developed in this paper allows to couple the pseudoscalar glueball to baryons.

We first consider the two-flavor case. The coupling of the pseudoscalar glueball $\tilde{G}$ to pseudoscalar mesons reads [61]:

$$\mathcal{L}_{\tilde{G}\Phi}^{N_f=2} = g_{\tilde{G}\Phi}^{N_f=2}\tilde{G}(\det\Phi - \det\Phi^\dagger)\;,$$

where $\Phi$ is the $2\times 2$ matrix given in Eq. (10). The coupling constant $g_{\tilde{G}\Phi}^{N_f=2}$ has dimension [energy]. The cou-pling to the glueball is obtained from Eq. (13) by replacing the determinant term with $\tilde{G}$ [for a preliminary discussion of this coupling, see Ref. [61]]:

$$\begin{aligned}\mathcal{L}_{\tilde{G}\Phi}^{N_f=2} &= ig_{\tilde{G}\Phi}^{N_f=2}\tilde{G}(\bar{\Psi}_2\Psi_1 - \bar{\Psi}_1\Psi_2)\\ &= -i\frac{g_{\tilde{G}\Phi}^{N_f=2}}{\cosh\delta}\tilde{G}(\bar{N}\gamma_5 N + \bar{N}_*\gamma_5 N_*\\ &\quad - \sinh\delta\,\bar{N}N_* + \sinh\delta\,\bar{N}_*N)\,.\end{aligned} \quad (26)$$

In the limit of zero mixing ($\delta \to \infty$), $\Psi_1 = N$ and $\Psi_2 = -N_*$. Hence, the interaction shows that a strong coupling of $\tilde{G}$ to $NN_*$ is realized. When mixing is present ($\delta < \infty$), one obtains

$$\Gamma_{\tilde{G}\to\bar{N}N} = \frac{(g_{\tilde{G}\Phi}^{N_f=2})^2 p_f^{NN}}{4\pi\cosh^2\delta}\,,$$

$$\begin{aligned}\Gamma_{\tilde{G}\to\bar{N}N_*+\text{h.c.}} &= \frac{(g_{\tilde{G}\Phi}^{N_f=2})^2\tanh^2\delta}{4\pi M_{\tilde{G}}^2}\\ &\quad \times [M_{\tilde{G}}^2 - (m_N + m_{N_*})^2] p_f^{N_*N}\,,\end{aligned}$$

where

$$p_f^{NN} = \sqrt{\frac{M_{\tilde{G}}^2}{4} - m_N^2}$$

and

$$p_f^{N_*N} = \frac{1}{2M_{\tilde{G}}}\sqrt{(M_{\tilde{G}}^2 - m_{N_*}^2 - m_N^2)^2 - 4m_{N_*}^2 m_N^2}$$

are the absolute values of the momenta of the final particles. Hence the ratio reads:

$$\frac{\Gamma_{\tilde{G}\to\bar{N}N}}{\Gamma_{\tilde{G}\to\bar{N}_*N+\text{h.c.}}} = \frac{M_{\tilde{G}}^2}{2\sinh^2\delta[M_{\tilde{G}}^2 - (m_N + m_{N_*})^2]}\frac{p_f^{NN}}{p_f^{N_*N}}\,.$$

Using Eq. (9) and the value $m_0 = (460\pm 136)$ MeV from Ref. [13] as well as the pseudoscalar glueball mass of $M_{\tilde{G}} = 2.6$ GeV [56], the numerical value of this ratio is

$$\frac{\Gamma_{\tilde{G}\to\bar{N}N}}{\Gamma_{\tilde{G}\to\bar{N}_*N+\text{h.c.}}} \simeq 1.96\,.$$

Thus, $\tilde{G} \to \bar{N}N$ is only slightly larger than $\tilde{G} \to \bar{N}_*N$, even if it has a much larger phase space. Neglecting phase space, one has:

$$\frac{M_{\tilde{G}}^2}{2\sinh^2\delta[M_{\tilde{G}}^2 - (m_N + m_{N_*})^2]} \simeq 0.85\,,$$

which shows that the coupling of $\tilde{G}$ to $\bar{N}_*N$ is expected to be sizable. While the numerical value depends on $m_0$ and is therefore model-dependent, a strong decay $\tilde{G} \to \bar{N}_*N$ can be viewed as a rather solid prediction. For this reason, it seems promising to search for the pseudoscalar glueball in the process

$$p + \bar{p} \to p + \bar{p}(1535) + \text{h.c.}\,. \quad (27)$$

at the future PANDA experiment [27].

Let us now turn to the generalization to $N_f = 3$. The coupling to ordinary mesons has the same formal expression as in Eq. (26),

$$\mathcal{L}_{\tilde{G}\Phi}^{N_f=3} = g_{\tilde{G}\Phi}^{N_f=3}\tilde{G}(\det\Phi - \det\Phi^\dagger)\,,$$

but now $\Phi$ is the $3\times 3$ matrix of Eq. (14) and $g_{\tilde{G}\Phi}^{N_f=3} = g_{\tilde{G}\Phi}^{N_f=2}/\phi_S$ is dimensionless. This Lagrangian was studied in detail in Ref. [61]. Concerning baryons, we follow the same procedure by replacing $\det\Phi - \det\Phi^\dagger$ with the glueball field $\tilde{G}$ in Eq. (22), obtaining the $U(3)_L \times U(3)_R$-symmetric expression

$$\begin{aligned}\mathcal{L}_{\tilde{G}B}^{N_f=3} &= i\frac{\tilde{g}_1 + \tilde{g}_2}{2}\tilde{G}\,\text{Tr}(\bar{B}_{M_*}B_N - \bar{B}_N B_{M_*}\\ &\quad - \bar{B}_{N_*}B_M + \bar{B}_M B_{N_*})\\ &\quad - i\frac{\tilde{g}_1 - \tilde{g}_2}{2}\tilde{G}\,\text{Tr}(\bar{B}_N\gamma_5 B_M + \bar{B}_M\gamma_5 B_N\\ &\quad + \bar{B}_{N_*}\gamma_5 B_{M_*} + \bar{B}_{M_*}\gamma_5 B_{N_*})\,.\end{aligned}$$

From this expression, one expects a strong coupling of $\tilde{G}$ to $\bar{B}_{M_*}B_N$, $\bar{B}_M B_{N_*}$, $\bar{B}_N\gamma_5 B_M$, and $\bar{B}_{M_*}\gamma_5 B_{N_*}$. By taking into account the pseudoscalar glueball mass of 2.6 GeV [56], there are only a few kinematically allowed decays. Summarizing, one expects the following sizable decays:

$$\tilde{G} \to N(1440)N\,,$$
$$\tilde{G} \to N(1535)N\,.$$

(The latter agrees with the $N_f = 2$ case, as it should.) When mixing among baryons is considered, also the decay $\tilde{G} \to \bar{N}N$ emerges for $N_f = 3$. While this coupling is important because it also induces the production of the glueball in proton-antiproton scattering, the corresponding amplitude should be smaller than $\tilde{G} \to N(1535)N$.

In conclusion, the present status of the search for the pseudoscalar glueball is still uncertain. However, decays into baryons seem to be potentially promising candidates to look for this elusive state. Also, the production of the pseudoscalar glueball in proton-antiproton scattering, see Eq. (27), is expected to be relevant.

Concerning the practical search of the pseudoscalar glueball at PANDA (and at other experiments), much depends on the phenomenological properties of the glueball and nearby states. Assuming that the lattice-QCD estimate is correct (i.e., the glueball $J^{PC} = 0^{-+}$ has a mass of about 2.6 GeV), a crucial question is the coupling strength of the glueball to mesons [which could not be evaluated in Refs. [60, 62]]. If the glueball turns out to be relatively narrow [as suggested by large-$N_c$ arguments, see the recent discussion in Ref. [64]], its experimental discovery will be easier. Another important feature is the existence of nearby pseudoscalar-isoscalar mesons (i.e., particles of the $\eta$-type): in the most favorable scenario, there is only one state whose decays are compatible with those of the pseudoscalar glueball. If other $\bar{q}q$ states are present, one should perform a more detailed study of mixing of the pseudoscalar-isoscalar sector. Undoubtedly, in the latter scenario the identification of the pseudoscalar glueball would be more difficult.

## V. SUMMARY AND CONCLUSIONS

A simple model which features only flavor symmetry cannot describe the decays $N(1535) \to N\eta$ and $\Lambda(1670) \to \Lambda\eta$; the numerical values for these decays are underestimated. On the contrary, $N(1650) \to N\eta$ and $\Lambda(1800) \to \Lambda\eta$ are in agreement with flavor symmetry. We have argued that, in the context of the mirror assignment for baryons, one can naturally add a term that embodies the axial anomaly in the baryonic sector. This term induces an additional interaction of the chiral partners with ground-state baryons and the $\eta$ meson. We have first discussed the consequences of this idea for the simpler two-flavor case and then extended it to the three-flavor case. In the latter, one can show that, after fixing the decay width for $N(1535) \to N\eta$, the decay $\Lambda(1670) \to \Lambda\eta$ can be also correctly described. Another result of our approach is a strong $N(1535)N\eta'$ coupling.

Finally, we studied the coupling of a putative pseudoscalar glueball to baryons by using the fact that the mathematical structure of this coupling resembles very closely that of the axial anomaly. We have found that the pseudoscalar glueball couples strongly to $N(1535)N$ and possibly to $N(1440)N$. It is therefore expected that the pseudoscalar glueball can be seen in the future PANDA experiment [27] by studying the process $p + \bar{p} \to p + \bar{p}(1535) + \text{h.c.}$.

As an outlook of the present work, we plan to study the $N_f = 3$ eLSM in full detail by evaluating the mixing among the four different baryonic octets. The anomaly studied in this work represents an important aspect of this investigation.

### Acknowledgments

We thank S. Gallas, G. Pagliara, P. Lakaschus, P. Kovacs, Gy. Wolf, A. Koenigstein, and R. Pisarski for useful discussions. LO acknowledges support by HGS-HIRe/HQM. MZ was supported by the Hungarian OTKA Fund No. K109462 and HIC for FAIR. FG acknowledges support from the Polish National Science Centre NCN through the OPUS project no. 2015/17/B/ST2/01625. DHR is supported in part by the High-End Visiting Expert project GDW20167100136 of the State Administration of Foreign Experts Affairs (SAFEA). FG and DHR acknowledge support from the DFG under grant no. RI 1181/6-1.

### APPENDIX A: DECAY WIDTHS OF THE (IMPROVED) FLAVOR MODEL

In this appendix we present the full expression of the (improved) flavor Lagrangian of Eq. (25), $\mathcal{L}_V^{\text{improved}} = \mathcal{L}_V + \mathcal{L}_A^{N_f=3}$. This Lagrangian contains the original flavor model $\mathcal{L}_V$ introduced in Sec. II, see Eq. (4), as well as the influence of the anomaly contained in $\mathcal{L}_A^{N_f=3}$ discussed in Sec. III B, see Eq. (24). After evaluating the traces, we extract the following terms describing decays into $\pi$, $K$, $\eta$, and $\eta'$:

$$\begin{aligned}
\mathcal{L}_V^{\text{improved}} =& \frac{i\lambda_V}{2\sqrt{3}}\bar{\Lambda}(\boldsymbol{\pi}\cdot\boldsymbol{\Sigma}_*) + \frac{i\lambda_V}{2\sqrt{3}}(\bar{\boldsymbol{\Sigma}}\cdot\boldsymbol{\pi})\Lambda_* + \frac{i\lambda_V}{2}\bar{\boldsymbol{\Sigma}}\cdot(\boldsymbol{\pi}\times\boldsymbol{\Sigma}_*) + \frac{i\lambda_V}{2}\bar{\boldsymbol{N}}(\boldsymbol{\pi}\cdot\boldsymbol{\tau})\boldsymbol{N}_* \\
&+ \frac{i\lambda_V}{6}\bar{\Lambda}(\eta_N + i2\sqrt{2}\eta_S)\Lambda_* + \frac{i\lambda_V}{2}\bar{\boldsymbol{\Sigma}}\eta_N \boldsymbol{\Sigma}_* + \frac{i\lambda_V}{4\sqrt{2}}\bar{\Xi}\eta_S\Xi_* + \frac{i\lambda_V}{2}\bar{\boldsymbol{N}}\eta_N\boldsymbol{N}_* \\
&- \frac{i\lambda_V}{\sqrt{3}}\bar{\Lambda}\boldsymbol{K}\cdot\boldsymbol{\Xi}_* + \frac{i\lambda_V}{2\sqrt{3}}\bar{\boldsymbol{N}}\cdot\boldsymbol{K}\Lambda_* - \frac{i\lambda_V}{2}\boldsymbol{K}^T(\bar{\boldsymbol{\Sigma}}\cdot\boldsymbol{\tau})\boldsymbol{\Xi}_* \\
&+ \frac{i\lambda_V}{\sqrt{3}}\bar{\Lambda}(\bar{\boldsymbol{K}}\cdot\boldsymbol{N}_*) - \frac{i\lambda_V}{2\sqrt{3}}\bar{\Xi}\cdot\bar{\boldsymbol{K}}\Lambda_* - \frac{i\lambda_V}{2}\bar{\Xi}(\boldsymbol{\Sigma}_*\cdot\boldsymbol{\tau})\bar{\boldsymbol{K}} \\
&+ \frac{i\tilde{\lambda}_A}{2\sqrt{3}}(\sqrt{2}\eta_N + \eta_S)[\bar{\Lambda}\Lambda_* + \bar{\boldsymbol{\Sigma}}\boldsymbol{\Sigma}_* + \bar{\Xi}\Xi_* + \bar{\boldsymbol{N}}\boldsymbol{N}_*] + \text{h.c.} \ .
\end{aligned} \quad \text{(A1)}$$

Because of the existing experimental data [1], we are especially interested in the decays of excited baryon resonances into ground-state baryons and a pseudoscalar meson, $\pi$, $\eta$, or $\bar{K}$. The Lagrangian describing the decay of a resonance $B_*$ into a ground-state baryon $B$ and a pseudoscalar meson $P = \pi, \eta, \bar{K}$ has the general struc-

TABLE IV: Coupling constants $g_{PBB_*}$ and $\gamma^{PBB_*}$ factors accounting for isospin.

| Decay | $g_{PBB_*}$ | $\gamma^{PBB_*}$ |
|---|---|---|
| $\Sigma_* \to \Lambda\pi$ | $\frac{i\lambda_V}{2\sqrt{3}}$ | 1 |
| $\Lambda_* \to \Sigma\pi$ | $\frac{i\lambda_V}{2\sqrt{3}}$ | 3 |
| $\Sigma_* \to \Sigma\pi$ | $\frac{i\lambda_V}{2}$ | 2 |
| $N_* \to N\pi$ | $\frac{i\lambda_V}{2}$ | 3 |
| $\Lambda_* \to \Lambda\eta$ | $\frac{i\lambda_V}{6}\cos\theta_P + \frac{i\tilde{\lambda}_A}{\sqrt{6}}(\sqrt{2}\cos\theta_P + \sin\theta_P)$ | 1 |
| $\Sigma_* \to \Sigma\eta$ | $\frac{i\lambda_V}{2}\cos\theta_P + \frac{i\tilde{\lambda}_A}{\sqrt{6}}(\sqrt{2}\cos\theta_P + \sin\theta_P)$ | 1 |
| $\Xi_* \to \Xi\eta$ | $\frac{i\lambda_V}{4\sqrt{2}}\cos\theta_P + \frac{i\tilde{\lambda}_A}{\sqrt{6}}(\sqrt{2}\cos\theta_P + \sin\theta_P)$ | 1 |
| $N_* \to N\eta$ | $\frac{i\lambda_V}{2}\cos\theta_P + \frac{i\tilde{\lambda}_A}{\sqrt{6}}(\sqrt{2}\cos\theta_P + \sin\theta_P)$ | 1 |
| $\Xi_* \to \Lambda\bar{K}$ | $-\frac{i2\lambda_V}{2\sqrt{3}}$ | 1 |
| $\Lambda_* \to N\bar{K}$ | $\frac{i\lambda_V}{2\sqrt{3}}$ | 1 |
| $\Sigma_* \to N\bar{K}$ | $\frac{i\lambda_V}{2}$ | 2 |

ture

$$\mathcal{L} = ig_{PBB_*}\bar{B}PB_* \,,$$

where the explicit expressions for the coupling constants $g_{PBB_*}$ can be obtained from the respective terms of the Lagrangian (A1) and are listed in Tab. IV. The respective tree-level decay widths can be calculated as

$$\Gamma_{B_* \to BP} = \gamma^{PBB_*} \frac{p_f}{8\pi m_{B_*}^2} \overline{|i\mathcal{M}_{B_* \to BP}|}^2$$
$$= \gamma^{PBB_*} \frac{p_f}{m_{B_*}} \frac{g_{PBB_*}^2}{4\pi}(E_B + m_B) \,,$$

where $E_B$ is the baryon energy in the rest frame of the decaying $B_*$, while the magnitude of the three-momenta of the decay products is

$$p_f = \frac{1}{2m_{B_*}}\sqrt{(m_{B_*}^2 - m_B^2 - m_P^2)^2 - 4m_B^2 m_P^2} \,.$$

The factor $\gamma^{PBB_*}$ takes into account the isospin of the involved particles. The respective values are listed in Tab. IV.

In the case $P = \eta$ one has to remember that the physical $\eta$ is a mixture of the pseudoscalar octet and the singlet, see Sec. II, Eq. (2). In this paper we have chosen $\theta_P = -44.6°$ obtained from Ref. [2]. The amplitude for a decay involving $\eta$ is given by

$$\mathcal{M}_{B_* \to B\eta} = \cos\theta_P \mathcal{M}_{B_* \to B\eta_N} + \sin\theta_P \mathcal{M}_{B_* \to B\eta_S} \,.$$

In this case the coupling constant $g_{\eta BB_*}$ is defined as a mixture of the constants of the non-strange and the strange sector:

$$g_{\eta BB_*} = g_{\eta_N BB_*}\cos\theta_P + g_{\eta_S BB_*}\sin\theta_P \,.$$

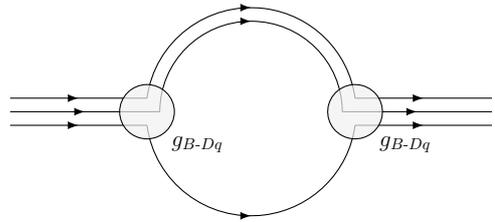

FIG. 1: Mass energy of the baryon.

The results in Tables I and II were obtained by setting the effects of the anomaly to zero, $\tilde{\lambda}_A$, while in Sec. III B the effects of $\tilde{\lambda}_A$ on decays and couplings have been studied.

## APPENDIX B: LARGE-$N_c$ SCALING PROPERTIES

In the large-$N_c$ limit, baryons are formed by $N_c$ quarks and their mass grows with $N_c$ [16], $m_B \propto N_c$. Indeed, its color wave function can be expressed as

$$B \equiv \varepsilon^{a_1 a_2 \dots a_{N_c}} q^{a_1} q^{a_2} \dots q^{a_{N_c}}$$

with $a_k = 1, 2, \dots, N_c$. In line with our approach, we may present a baryon in the large-$N_c$ limit as an $(N_c - 1)$-quark and a quark. An $(N_c - 1)$-quark is a generalization of the diquark:

$$D^{a_1} \equiv \varepsilon^{a_1 a_2 \dots a_{N_c}} q^{a_2} \dots q^{a_{N_c}} \,.$$

As a consequence also the mass of a generalized diquark scales as $N_c$, $m_D \propto N_c$. Moreover, the baryon can still be expressed as a 'generalized diquark'-quark object:

$$B \equiv D^{a_1} q^{a_1} \,.$$

The basic diagram is the mass energy of a baryon, see Fig. 1. Since the baryon mass must grow with $N_c$, the coupling constant $g_{B-Dq}$ scales as $\sqrt{N_c}$, in such a way that the whole amplitude corresponding to Fig. 1 contributes as $\left(g_{B-Dq}\frac{1}{m_D}g_{B-Dq}\right)N_c \propto N_c$ (where the overall factor $N_c$ arises from the circulating color $a_1$).

The dominant interaction of a baryon with a meson is depicted in Fig. 2 and scales as

$$\left(g_{B-Dq}\frac{1}{m_D}g_{M-\bar{q}q}g_{B-Dq}\right)N_c \propto \sqrt{N_c} \,,$$

where we have used that the coupling of a standard meson to $\bar{q}q$ scales $\sim g_{M-\bar{q}q} \propto 1/\sqrt{N_c}$, in agreement with Ref. [16]. Figure 2 corresponds to the dominant term in our interaction Lagrangian of Eq. (4). The terms in the parenthesis correspond to the various elements of the diagram; the generalized diquark, being heavy, contributes as $1/m_D \propto 1/N_c$. An overall factor $N_c$ emerges for the same reasons as in Fig. 1.

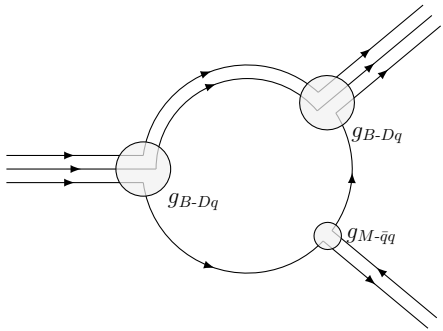

FIG. 2: Dominant baryon-meson interaction term.

We now show that this term corresponds to the term of Eq. (4). By rendering the flavor indices in the flavor traces explicit [see Ref. [14] for details], and remembering that a baryon is a diquark-quark object ($O_{ij} \equiv D_j q_i$) and a meson a quark-antiquark object ($P_{ij} \equiv \bar{q}_j q_i$), the first term of Eq. (4) is:

$$i\lambda_V \mathrm{Tr}[\bar{O}PO^*] = i\lambda_V \bar{O}_{ij} P_{jk} O_{ki}^* \\ \equiv i\lambda_V \left(\bar{q}_j \bar{D}_i\right) \left(\bar{q}_k q_j\right) \left(D_i q_k\right) .$$

It is evident that the quark lines of the diquark field are closed (same index $i$), while the quark lines of the in- and outgoing baryons are linked to the produced meson, just as Fig. 2 shows.

Next, we consider the case where the outgoing meson emerges from (at least) two gluons forming a white configuration. The corresponding diagrams are presented in Fig. 3 (the second diagram uses the double-quark line notation for the gluons and helps to clarify the $N_c$ counting). The large-$N_c$ scaling of this contribution is

$$\left(g_{B\text{-}Dq} \frac{1}{m_D} g_{M\text{-}\bar{q}q} g_{QCD}^4 g_{B\text{-}Dq}\right) N_c^2 \propto \frac{1}{\sqrt{N_c}} ,$$

where one factor of $N_c$ arises from the color circulating in the main part of the diagram (as in Figs. 1 and 2) and another one from the color circulating in the loop created by the di-gluon exchange. The whole diagram is hence suppressed by a factor $N_c$ with respect to the dominant term. This interaction corresponds to the term proportional to $\gamma_V$ in Eq. (5). Namely:

$$i\gamma_V \mathrm{Tr}[\bar{O}O^*]\mathrm{Tr}[P] = i\gamma_V \bar{O}_{ij} O_{ji}^* P_{kk} \\ \equiv i\gamma_V \left(\bar{q}_j \bar{D}_i\right) \left(D_i q_j\right) \left(\bar{q}_k q_k\right) ,$$

where it is clear that the flavor of the external meson is not exchanged with the quark of the baryon. This term generates also a coupling of the type $\bar{N}^* N \eta_S$, which is not possible for the dominant term proportional to $\lambda_V$. Namely, a strange-antistrange pair can be attached to a baryon-baryon coupling only via a di-gluon in a white configuration.

In the pseudoscalar channel, one can intuitively understand this term via the exchange of two gluons in the pseudoscalar configuration, whose coupling is $\left(\bar{q}_j \bar{D}_i\right)(D_i q_j) G^a_{\mu\nu} \tilde{G}^{a,\mu\nu}$. Then, $G^a_{\mu\nu} \tilde{G}^{a,\mu\nu}$ couples to the singlet configuration $\eta_0 \equiv \bar{q}^b_k q^b_k$ via $\eta_0 \, G^a_{\mu\nu} \tilde{G}^{a,\mu\nu}$. The quantity $G^a_{\mu\nu} \tilde{G}^{a,\mu\nu}$ is related to the QCD axial anomaly, which increases the amplitude for the production of $\eta_0$, as described at length in the main text.

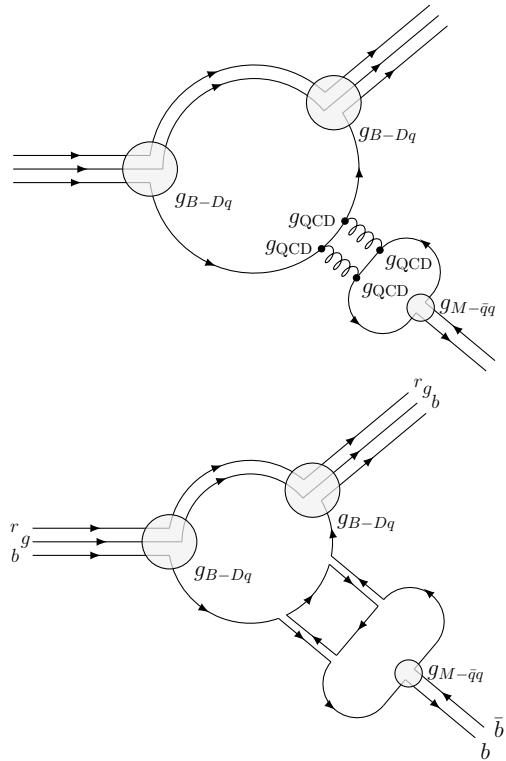

FIG. 3: Baryon-meson interaction with intermediate gluons, where the first diagram represents the flavor flow, and the second one represents the color flow.

The last term that needs to be studied is $i\beta_V \mathrm{Tr}[\bar{O}O^* P]$. Contrary to naive expectations, this is the most subtle and difficult one. The flavor structure reads

$$i\beta_V \mathrm{Tr}[\bar{O}O^* P] = i\beta_V \bar{O}_{ij} O^*_{jk} P_{ki} \\ \equiv i\beta_V \left(\bar{q}_j \bar{D}_i\right)(D_k q_j)\left(\bar{q}_i q_k\right) .$$

Here, one observes that the quark-antiquark pair of the meson couples to the diquark, while the single quark line with flavor $j$ goes through undisturbed. Recalling that $D_i = \varepsilon_{imn} q_m q_n$, various terms exist. For instance, for $i = 3$ and $k = 3$ the coupling term $\bar{D}_3 D_3 \left(\bar{q}_3 q_3\right)$ implies the transition $[u,d] \to [u,d]\bar{s}s$, which shows the emergence of an $\bar{s}s$ pair from a non-strange structure, hence this term must be large-$N_c$ suppressed. On the other hand, for $i = 2$ and $k = 3$ one has $\bar{D}_2 D_3 \left(\bar{q}_2 q_3\right)$, which implies $[u,s] \to [u,d]\bar{d}s$, therefore at first glance simple flavor-connected diagrams seem to be possible.

However, when color is taken into account the situation is not that simple. Restoring the color indices, the inter-

action term reads

$$i\beta_V \left(\bar{q}_j^a \bar{D}_i^a\right) \left(D_k^a q_j^a\right) \left(\bar{q}_i^a q_k^a\right) , \quad (B1)$$

which means that transitions of the type $[R,G] \to [R,G]\bar{B}B$ occur (for $a = 3$, $N_c = 3$ and $D^a = \varepsilon^{abc}q^b q^c$), i.e., color changes. Note that we keep only one color index for simplicity; other diagrams with different color lines exist, but they are either of the same order in large $N_c$ (hence, can be formally reabsorbed in the hadron-quark vertices) or they are further suppressed. The color index being the same for all objects in Eq. (B1) and in virtue of the Levi-Civita tensor defining the diquarks, *only* those transitions are allowed in which the emerging color-anticolor is different from the one carried by the quarks of the diquark:

$$\bar{D}_i^a D_k^a \left(\bar{q}_i^a q_k^a\right) = \left[\varepsilon^{abc}\varepsilon_{irs}\bar{q}_r^b \bar{q}_s^c\right] \left[\varepsilon^{ab'c'}\varepsilon_{kr's'}q_{r'}^{b'} q_{s'}^{c'}\right] \left(\bar{q}_i^a q_k^a\right) .$$

In the case $i=2$ and $k=3$ as mentioned above, one has (upon setting $a = 3$) $\bar{D}_2^3 D_3^3 \bar{q}_2^3 q_3^3$, hence a transition of the type $s_R u_G \to u_R d_G (\bar{d}_B s_B)$ follows. This transition is not possible by simply exchanging quark lines, but additional gluons that properly switch color are necessary. Instead of searching for these gluon configurations, we use an alternative elegant way to achieve these transitions by taking into account all the features listed above: we make use of an *additional* white intermediate virtual baryon, as depicted in Fig. 4. Namely, in this way one *automatically* couples a diquark to a quark with the right 'missing' color. Moreover, the large-$N_c$ scaling of this can be easily calculated by using the previously introduced scaling properties:

$$\left(g_{B\text{-}Dq} \frac{1}{m_D} g_{B\text{-}Dq} g_{M\text{-}\bar{q}q} \frac{1}{m_B} g_{B\text{-}Dq} \frac{1}{m_D} g_{B\text{-}Dq}\right) N_c$$
$$\propto N_c^{-1/2} .$$

In the end, this term is also suppressed with $N_c$ and $\beta_V$ scales as $1/\sqrt{N_c}$, just as $\gamma_V$.

The above considerations can be carried out for arbitrary $N_c$ in a straightforward way provided that the number of flavors $N_f$ equals $N_c$. Namely, in this way the generalized diquark reads $D_k^a = \varepsilon^{aa_2...a_{N_c}}\varepsilon^{kk_2...k_{N_c}}q^{i_2,k_2}...q^{ik_{N_c}}$ and all the traces above are defined in a way similar to the physical case $N_f = N_c = 3$. Extensions to $N_f \neq N_c$, are possible, but would require a more detailed study, which goes beyond the scope of the present work. Here, large-$N_c$ arguments are needed to distinguish dominant and subdominant terms.

### APPENDIX C: EXPLICIT EXPRESSIONS FOR THE DECAY WIDTHS IN THE ELSM FOR $N_f = 2$

The inclusion of the anomaly term (13) into the model of Ref. [13] yields an additional contribution to the decay width of $N_* \to N\eta$:

$$\Gamma_{N_* \to N\eta} = \Gamma_{N_* \to N\eta}^{\text{without anomaly}} + \lambda_\eta \frac{p_f}{2\pi} \frac{m_N}{m_{N_*}} \frac{Z^2}{2} \lambda_A \phi_N \left\{ \left[-\frac{1}{2}(\hat{g}_1 - \hat{g}_2)\frac{\sinh\delta}{\cosh^2\delta} + \lambda_A \phi_N \tanh^2\delta\right] \left(\frac{E_N}{m_N} + 1\right) \right.$$
$$\left. - \frac{1}{2}w(c_1 + c_2)\frac{\sinh\delta}{\cosh^2\delta} \left(\frac{m_{N_*}^2 - m_N^2 - m_\eta^2}{2m_N} + E_\eta\right) \right\} ,$$

with

$$\Gamma_{N_* \to N\eta}^{\text{without anomaly}} = \lambda_\eta \frac{p_f}{2\pi} \frac{m_N}{m_{N_*}} \frac{Z^2}{32\cosh^2\delta} \left\{ (\hat{g}_1 - \hat{g}_2)^2 \left(\frac{E_N}{m_N} + 1\right) \right.$$
$$+ w^2(c_1 + c_2)^2 \left[(m_{N_*}^2 - m_N^2 - m_\eta^2)\frac{E_\eta}{m_N} + m_\eta^2\left(1 - \frac{E_N}{m_N}\right)\right]$$
$$\left. + 2(\hat{g}_1 - \hat{g}_2)w(c_1 + c_2)\left(\frac{m_{N_*}^2 - m_N^2 - m_\eta^2}{2m_N} + E_\eta\right) \right\} ,$$

where the constants $\hat{g}_1 = 10.2 \pm 0.7$, $\hat{g}_2 = 17.3 \pm 0.8$, $c_1 = -2.65 \pm 0.18$, $c_2 = 10.2 \pm 2.6$, $Z = 1.81$ were obtained in Ref. [21] and parametrize the interactions of baryonic fields with scalar and pseudoscalar mesons. The factor $\lambda_\eta = \cos^2\theta_P$ [$\theta_P = -40°$ in Ref. [21]] takes into account the mixing of Eq. (2), where it is assumed that the amplitude of the decay $N_* \to N\eta_S$ is suppressed [this is not in agreement with the anomaly term studied here; this is why in Ref. [13] a too small decay $N_* \to N\eta$ was obtained]. Furthermore, we introduced the energy of the nucleon and the $\eta$ meson together with the modulus of

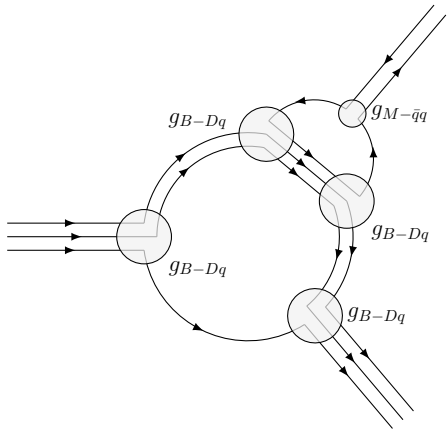

FIG. 4: Baryon-meson interaction with an intermediate virtual baryon.

the three-momentum of the two outgoing particles:

$$E_N = \sqrt{p_f^2 + m_N^2}\ , \quad E_\eta = \sqrt{p_f^2 + m_\eta^2}\ ,$$

with

$$p_f = \frac{1}{2m_{N_*}}\sqrt{(m_{N_*}^2 - m_N^2 - m_\eta^2)^2 - 4m_N^2 m_\eta^2}\ .$$

The new constant $\lambda_A$ influences only the $\eta$ decay and therefore it can be chosen such that it correctly describes the decay width of $N_* \to N\eta$. Choosing

$$\lambda_A^{N_f=2} = 0.006\ \text{MeV}^{-1} \quad \text{or} \quad \lambda_A^{N_f=2} = -0.011\ \text{MeV}^{-1}$$

allows for a correct description of the decay $N(1535) \to N\eta$.

### APPENDIX D: ADDITIONAL DETAILS FOR THE ANOMALY TERM FOR $N_f = 3$

We now turn in more detail to the interaction with baryons discussed in Sec. III B. To this end, we simplify the notation by introducing the vectors

$$\begin{aligned}\boldsymbol{X} &:= (X_N, \gamma_5 X_{N_*}, X_M, \gamma_5 X_{M^*})^T\ ,\\ \bar{\boldsymbol{X}} &:= (\bar{X}_N, -\bar{X}_{N_*}\gamma_5, \bar{X}_M, -\bar{X}_{M^*}\gamma_5)\ .\end{aligned}$$

They combine the four types of particles or resonances which are included in the baryon octet, i.e., $X = N, \Lambda, \Sigma$, or $\Xi$. In this way, we can write the anomaly Lagrangian in a very compact form:

$$\mathcal{L}_A^{N_f=3} = -i\bar{\boldsymbol{X}}\gamma_5 \hat{\eta}_N^A \eta_N \boldsymbol{X} - i\bar{\boldsymbol{X}}\gamma_5 \hat{\eta}_S^A \eta_S \boldsymbol{X}\ + \ldots\ ,$$

where we omitted all terms describing four- and five-point interactions. In this expression $\hat{\eta}_N^A$ and $\hat{\eta}_S^A$ are $4 \times 4$ matrices containing the coupling constants:

$$\hat{\eta}_N^A = \frac{Z_{\eta_N} \phi_N \phi_S}{2\sqrt{2}} \begin{pmatrix} 0 & 0 & \lambda_{A1} - \lambda_{A2} & \lambda_{A1} + \lambda_{A2} \\ 0 & 0 & -(\lambda_{A1} + \lambda_{A2}) & -(\lambda_{A1} - \lambda_{A2}) \\ \lambda_{A1} - \lambda_{A2} & -(\lambda_{A1} + \lambda_{A2}) & 0 & 0 \\ \lambda_{A1} + \lambda_{A2} & -(\lambda_{A1} - \lambda_{A2}) & 0 & 0 \end{pmatrix}\ ,$$

$$\hat{\eta}_S^A = \frac{Z_{\eta_S} \phi_N^2}{4\sqrt{2}} \begin{pmatrix} 0 & 0 & \lambda_{A1} - \lambda_{A2} & \lambda_{A1} + \lambda_{A2} \\ 0 & 0 & -(\lambda_{A1} + \lambda_{A2}) & -(\lambda_{A1} - \lambda_{A2}) \\ \lambda_{A1} - \lambda_{A2} & -(\lambda_{A1} + \lambda_{A2}) & 0 & 0 \\ \lambda_{A1} + \lambda_{A2} & -(\lambda_{A1} - \lambda_{A2}) & 0 & 0 \end{pmatrix}\ .$$

We see that the anomaly Lagrangian describes interactions of the $\eta$ meson with two baryons with different parity, indicated by the index combinations $(N, M_*)$ or $(M, N_*)$, or with equal parity, indicated by the index combination $(N, M)$ or $(N_*, M_*)$.

With the definition of the $\boldsymbol{X}$ vector of fields introduced above, also the interaction Lagrangian involving the pseudoscalar glueball (discussed in Sec. IV) can be rewritten in a compact form:

$$\mathcal{L}_{\tilde{G}B}^{N_f=3} = -\bar{\boldsymbol{X}}\gamma_5 \hat{\mathcal{G}} \tilde{G} \boldsymbol{X}\ ,$$

where the $4 \times 4$ matrix $\hat{\mathcal{G}}$ contains the coupling constants:

$$\hat{\mathcal{G}} = \frac{1}{2}\begin{pmatrix} 0 & 0 & \tilde{g}_1 - \tilde{g}_2 & \tilde{g}_1 + \tilde{g}_2 \\ 0 & 0 & -(\tilde{g}_1 + \tilde{g}_2) & -(\tilde{g}_1 - \tilde{g}_2) \\ \tilde{g}_1 - \tilde{g}_2 & -\tilde{g}_1 + \tilde{g}_2) & 0 & 0 \\ \tilde{g}_1 + \tilde{g}_2 & -(\tilde{g}_1 - \tilde{g}_2) & 0 & 0 \end{pmatrix}\ .$$

At present, the coupling constants $\tilde{g}_1$ and $\tilde{g}_2$ are unknown.